\documentclass[conference]{IEEEtran}
\IEEEoverridecommandlockouts
\usepackage{amsmath,graphicx}
\usepackage{booktabs}
\usepackage{balance}

\usepackage{amsmath,empheq, epsfig,amssymb,amsthm,url,amsfonts}
\usepackage{algorithm}% http://ctan.org/pkg/algorithm
\usepackage{algpseudocode}% http://ctan.org/pkg/algorithmicx
\usepackage{multirow}
\usepackage{mdframed}
\usepackage{url}
\usepackage{enumitem}
\usepackage[makeroom]{cancel}
\usepackage{dblfloatfix}
\usepackage{array}
\usepackage{comment}
\usepackage{epstopdf}
\usepackage{float}
\usepackage{subfig}
\usepackage{breqn}
\usepackage{cases}
\usepackage{caption}
\usepackage{tabularx}
\usepackage[printonlyused,withpage]{acronym}
\usepackage{balance}
\usepackage{graphbox} %loads graphicx package

\usepackage{hyperref}
\usepackage[table]{xcolor} % For shading rows in tables
\usepackage{colortbl}
\definecolor{gray}{rgb}{0.9,0.9,0.9} % Define a light gray color
\usepackage{tikz}
\usetikzlibrary{fit,calc}
\newcommand*{\tikzmk}[1]{\tikz[remember picture,overlay,] \node (#1) {};\ignorespaces}
\newcommand{\boxit}[1]{\tikz[remember picture,overlay]{\node[yshift=0.1pt,fill=#1,opacity=.25,fit={(A)($(B)+(.87\linewidth,0.1\baselineskip)$)}] {};}\ignorespaces}
\colorlet{mypink}{red!40}
\colorlet{myblue}{cyan!60}

\setlist[itemize]{label=$\triangleright$}
\newtheoremstyle{break}% name
{}%         Space above, empty = `usual value'
{}%         Space below
{\itshape}% Body font
{}%         Indent amount (empty = no indent, \parindent = para indent)
{\bfseries}% Thm head font
{.}%        Punctuation after thm head
{\newline}% Space after thm head: \newline = linebreak
{}%         Thm head spec
\theoremstyle{break}

\theoremstyle{definition}

\usepackage{caption}



\newcommand{\vect}[1]{\mathbf{#1}}
\newcommand{\bs}[1]{\boldsymbol{#1}}
\newcommand{\E}{\mathbb{E}}
\usepackage{url}

\makeatletter
\def\thmhead@plain#1#2#3{%
	\thmname{#1}\thmnumber{\@ifnotempty{#1}{ }\@upn{#2}}%
	\thmnote{ {\the\thm@notefont#3}}}
\let\thmhead\thmhead@plain
\makeatother

\graphicspath{{./figs/}}

\newcommand{\argmax}{\operatornamewithlimits{argmax}}
\newcommand{\argmin}{\operatornamewithlimits{argmin}}

\newcommand{\lk}{ \left\{ }
\newcommand{\rk}{ \right\} }

\newcommand{\diag}{\mbox{{diag}}}

\newsavebox\mybox

\acrodef{SE}{Speech enhancement}
\acrodef{AVSE}{audio-visual speech enhancement}
\acrodef{STFT}{short-time Fourier transform}
\acrodef{ESTOI}{extended short-time objective intelligibility}
\acrodef{NMF}{non-negative matrix factorization}
\acrodef{DNN}{deep neural network}
\acrodef{VAE}{variational auto-encoder}
\acrodef{DKF}{deep Kalman filter}
\acrodefplural{VAEs}{variational auto-encoders}
\acrodef{EM}{expectation-maximization}
\acrodef{TF}{time-frequency}
\acrodef{ELBO}{evidence lower bound}
\acrodef{LR}{Living Room}
\acrodef{SDR}{signal-to-distortion ratio}
\acrodef{PESQ}{perceptual evaluation of speech quality}
\acrodef{SNR}{signal-to-noise ratio}
\acrodef{DNNs}{deep neural networks}
\acrodef{VESDE}{variance-exploding stochastic differential equation}
\acrodef{SDE}{stochastic differential equation}
\acrodef{GAN}{generative adversarial networks}
\acrodefplural{GANs}{generative adversarial networks}
\acrodef{SI-SDR}{scale-invariant signal-to-distortion ratio}
\acrodef{MOS}{mean opinion score}
\acrodef{SGMSE+}{score-based generative model for speech enhancement}
\acrodef{NCSNPP++}{Noise-Conditional Score Network}
\acrodef{WSJ}{Wall Street Journal}
\acrodef{UDiffSE}{Unsupervised Diffusion-Based Speech Enhancement}
\acrodef{PC}{Predictor-Corrector}
\acrodef{DMPS}{Diffusion Model Posterior Sampling}
\acrodef{NN}{neural network}

\newcommand{\normalc}{\mathcal{N}_{\mathbb{C}}(\vect{0}, \vect{I})}

\usepackage{bm}

\usepackage{cite}
\usepackage{textcomp}
\def\BibTeX{{\rm B\kern-.05em{\sc i\kern-.025em b}\kern-.08em
    T\kern-.1667em\lower.7ex\hbox{E}\kern-.125emX}}
\begin{document}

\title{Diffusion-based Unsupervised Audio-visual Speech Enhancement

%\title{{\fontsize{13pt}{18pt}\selectfont \textbf{DIFFUSION-BASED UNSUPERVISED AUDIO-VISUAL SPEECH ENHANCEMENT}}

%\title{{\Large \textbf{DIFFUSION-BASED UNSUPERVISED AUDIO-VISUAL SPEECH ENHANCEMENT}}

\thanks{This work was supported by the French National Research Agency (ANR) under the project REAVISE (ANR-22-CE23-0026-01).}
}

% \author{\IEEEauthorblockN{Jean-Eudes Ayilo}
% \IEEEauthorblockA{\textit{Multispeech Team, Inria} \\
% Nancy, France }
% \and
% \IEEEauthorblockN{Mostafa Sadeghi}
% \IEEEauthorblockA{\textit{Multispeech Team, Inria} \\
% Nancy, France }
% \and
% \IEEEauthorblockN{Romain Serizel}
% \IEEEauthorblockA{\textit{Multispeech Team, Inria} \\
% Nancy, France }
% \and
% \IEEEauthorblockN{Xavier Alameda-Pineda}
% \IEEEauthorblockA{\textit{RobotLearn Team, Inria} \\
% Grenoble, France }
% }

% \author{\IEEEauthorblockN{Jean-Eudes Ayilo$^1$, Mostafa Sadeghi$^1$, Romain Serizel$^1$, Xavier Alameda-Pineda$^2$}
% \IEEEauthorblockA{
% \textit{1 Université de Lorraine, CNRS, Inria, LORIA, F-54000 Nancy \\
% \textit{2 Inria Grenoble \& Univ. Grenoble Alpes, France}
% }
% jean-eudes.ayilo@inria.fr, mostafa.sadeghi@inria.fr,  romain.serizel@loria.fr, xavier.alameda-pineda@inria.fr}
% }

\author{\IEEEauthorblockN{Jean-Eudes Ayilo\textsuperscript{1}, Mostafa Sadeghi\textsuperscript{1}, Romain Serizel\textsuperscript{1}, Xavier Alameda-Pineda\textsuperscript{2}}
\IEEEauthorblockA{\textit{\textsuperscript{1}Université de Lorraine, CNRS, Inria, Loria, Nancy, France}\\
\textit{\textsuperscript{2}Université Grenoble Alpes, Inria, Grenoble, France}\\
jean-eudes.ayilo@inria.fr, mostafa.sadeghi@inria.fr, romain.serizel@loria.fr, xavier.alameda-pineda@inria.fr}
}

% \name{%
% Jean-Eudes Ayilo$^1$, %
% Mostafa Sadeghi$^1$, %
% Romain Serizel$^1$, and
% Xavier Alameda-Pineda$^2$ \thanks{This work was supported by the French National Research Agency (ANR) under the project REAVISE (ANR-22-CE23-0026-01)}
% }
% \address{%
% $^1$Université de Lorraine, CNRS, Inria, LORIA, F-54000 Nancy, France\\ 
% $^2$Inria Grenoble \& Univ. Grenoble Alpes, France}

\maketitle

\begin{abstract}
This paper proposes a new unsupervised audio-visual speech enhancement (AVSE) approach that combines a diffusion-based audio-visual speech generative model with a non-negative matrix factorization (NMF) noise model. First, the diffusion model is pre-trained on clean speech conditioned on corresponding video data to simulate the speech generative distribution. This pre-trained model is then paired with the NMF-based noise model to estimate clean speech iteratively. Specifically, a diffusion-based posterior sampling approach is implemented within the reverse diffusion process, where after each iteration, a speech estimate is obtained and used to update the noise parameters. Experimental results confirm that the proposed AVSE approach not only outperforms its audio-only counterpart but also generalizes better than a recent supervised-generative AVSE method. 
%shows slight edge competitive results in mismatch conditions compared with a recent supervised-generative AVSE method.
Additionally, the new inference algorithm offers a better balance between inference speed and performance compared to the previous diffusion-based method. Code and demo available at: \url{https://jeaneudesayilo.github.io/fast_UdiffSE}
\end{abstract}


\section{Introduction}
\ac{SE} refers to the problem of extracting a clean speech signal from a noisy recording. Early algorithms implemented for solving this task relied solely on acoustic features. However, speech production is inherently multimodal, involving movements of the lips and tongue, for example. Research in speech perception has demonstrated a crucial impact of visual cues on the ability of humans to focus their auditory
attention on a speech signal {\cite{michelsanti2021overview, mcgurk1976hearing, sumby1954visual}. Consequently, \ac{AVSE} has emerged as a new research trend. In this approach, lip movements are primarily used as complementary information to acoustic features to improve the performance, particularly in low \ac{SNR} environments \cite{michelsanti2021overview, blanco2023avse}.

Existing \ac{DNN}-based AVSE frameworks (and SE in general) are primarily divided into two learning approaches: supervised \cite{alfouras2018conversation, hou2018audio, chuang2022improved, afouras2019my,  richter2023audio, jung2024flowavse,richter2023speech, pascual2017segan} and unsupervised ~\cite{bie2022unsupervised, %tzinis2022remixit, 
Sadeghi_2020, golmakani2023audio}. Supervised methods, whether predictive or generative, involve training a \ac{DNN} on pairs of clean and noisy speech, and possibly corresponding visual data. Specifically, predictive methods focus on mapping noisy speech directly to clean speech or to a time-frequency mask. In contrast, generative methods aim to generate clean speech at inference by learning the distribution of clean speech conditioned on the noisy input, rather than directly mapping from noisy to clean speech. 
%%%initial version: we restrict to sgmse uniquely
% A recent supervised-generative methodology for \ac{SE}  \cite{richter2023speech} leverages diffusion models \cite{song2021scorebased}, where clean speech training data are gradually degraded by both environmental and Gaussian noise in a forward process. This process transforms the clean speech distribution to a Gaussian distribution centered on noisy speech \cite{richter2023speech}. Then, in a reverse process, the clean speech is iteratively recovered from the input noisy speech via a learned \ac{DNN}. This approach has been extended to \ac{AVSE} by incorporating lips video features as additional conditional information \cite{richter2023audio}. 

%%%2nd version: here we enlarge a little bit to paper similar to sgmse
% Some recent supervised-generative methodologies for \ac{SE}  \cite{richter2023speech, lemercier2023storm, gonzalez2024diffusion} leverage diffusion models \cite{song2021scorebased}, where clean speech training data are gradually degraded by both environmental and Gaussian noise in a forward process. This process transforms the clean speech distribution to a Gaussian distribution centered on noisy speech. Then, in a reverse process, the clean speech is iteratively recovered from the input noisy speech via a learned \ac{DNN}. Lips video features can also be incorporated in such \ac{SE} models as shown by \cite{richter2023audio} to extend \cite{richter2023speech}. 

%%%3rd version: here we include all possible supervised-diffusion-based SE 
Some recent supervised-generative methodologies for \ac{SE}  \cite{richter2023speech, %lemercier2023storm,
gonzalez2024diffusion,serra2022universal, lu2022conditional,yen2023cold} leverage diffusion models \cite{song2021scorebased}. A diffusion model learns data distribution by first pushing the clean data distribution towards a prior Gaussian distribution via the progressive corruption of the clean data with Gaussian noise. A \ac{DNN} is trained to progressively transform samples drawn from the prior Gaussian distribution into clean data. In the diffusion-based supervised-generative SE context, the diffusion model which aims to learn clean speech distribution is conditioned on noisy speech. %, so that clean speech is recovered staring from noisy speech and gaussian noise. 
Lips video features can also be incorporated in such \ac{SE} models as shown by \cite{richter2023audio}. 
%%%suite of the intro
As diffusion model inference requires many iterations, a faster alternative to \cite{richter2023audio}, called FlowAVSE, was proposed in  \cite{jung2024flowavse}. It is a two-stage method that uses in its first stage a supervised-predictive network that outputs an estimate of clean speech given the noisy speech and the lips video. Then, a conditional generative flow matching algorithm generates the final enhanced speech in just one sampling step, conditionally to the lips video and the output of the first stage. Despite their generative nature, these methods still require pairs of clean and noisy speech for training. Although all these supervised methods %they 
can generalize to unseen noise conditions at test time, a significant amount of paired data is needed \cite{lin2023unsupervised, ec0b16359ea445e08c2672071f5542ac}. This is because it is impossible to train these models across all potential noise types and acoustic scenarios \cite{bie2022unsupervised}.
 
To enhance the robustness of SE models to unseen noises during training, some \textit{unsupervised} methods have been developed that do not require a noise dataset. Here, the training stage involves learning the prior distribution of clean speech using models like \ac{VAE} \cite{carbajal2021disentanglement, Sadeghi_2020} or dynamical \ac{VAE} \cite{golmakani2023audio}. This learned prior is then combined with a noise model based on \ac{NMF} \cite{lee2000algorithms} to estimate clean speech using \ac{EM}. In this vein, Nortier \textit{et al.} \cite{nortier2023unsupervised} recently introduced UDiffSE, an {unsupervised} audio-only \ac{SE} framework utilizing diffusion models. In this approach, an unconditional diffusion model is trained on clean speech data. This diffusion model, serving as a prior for clean speech, is then combined with an NMF-based noise model to enhance speech. This method has been shown to outperform \ac{VAE}-based \ac{SE} approaches on some metrics and provides better generalization than its supervised counterpart, i.e. \cite{richter2023speech}.

In this paper, we propose a diffusion-based \ac{AVSE} framework that learns a clean speech distribution from audio-visual (AV) data. The model incorporates visual features extracted from a self-supervised audio-visual speech representation learning model \cite{shi2022learning} as conditioning information. Additionally, we develop an iterative inference algorithm, named UDiffSE+, which operates significantly faster than UDiffSE. This approach substantially reduces the total number of iterations required, without dramatically degrading performance, as our experimental results will demonstrate.

%%%previously commented
%This considerably reduces the overall number of iterations, without significant performance degradation. Our experimental results showcase the improved performance of our proposed diffusion-based \ac{AVSE} framework over \ac{VAE}-based approaches.

%%%previously commented
% The remainder of this paper is structured as follows. In Section~\ref{sec:related}, we review the UDiffSE framework, setting the foundation for our work. The proposed AVSE methodology is introduced in Section~\ref{sec:proposed}, including the extension of UDiffSE to the audio-visual case and the introduction of a fast inference algorithm. Section~\ref{sec:exp} presents experimental results, and Section~\ref{sec:conc} concludes the paper.
%The remainder of this paper is structured as follows. In Section~\ref{sec:udiffse}, we review the UDiffSE framework, setting the foundation for our work. The proposed AVSE methodology is introduced in Section~\ref{sec:proposed}. Section~\ref{sec:exp} presents experimental results, and Section~\ref{sec:conc} concludes the paper.
\begin{figure*}[t]
 \centering
 \includegraphics[width=0.75\linewidth]{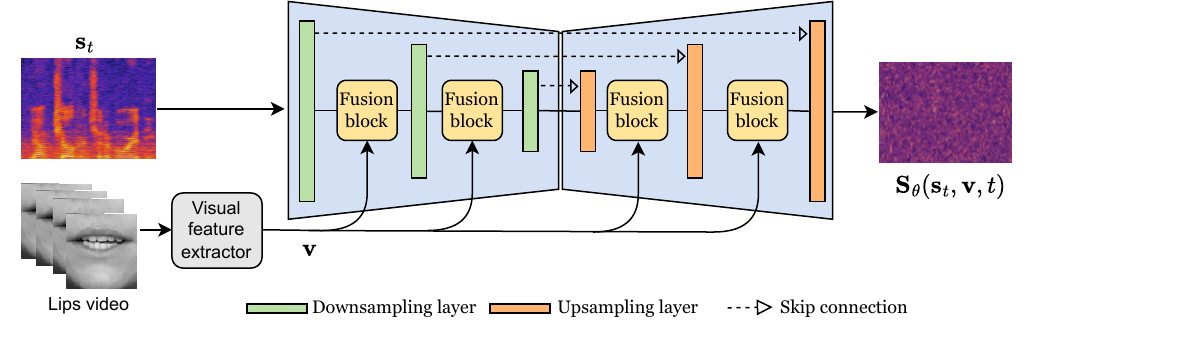}
 \caption{Schematic diagram of the proposed AV-U-Net (score model) architecture.}
 \label{fig:av_score}
\end{figure*}
\section{Diffusion-based unsupervised SE}\label{sec:udiffse}
In this section, we review the diffusion-based unsupervised (audio-only) \ac{SE} framework \cite{nortier2023unsupervised} i.e. UDiffSE. The modeling is done in the complex-valued \ac{STFT} domain. The observation model is $\vect{x} = \vect{s} + \vect{n}$, where $\vect{x}$, $\vect{s}$, and $\vect{n}$ denote \ac{STFT} arrays of noisy (mixture) speech, clean speech, and background noise, respectively. For notational simplicity, 2D \ac{STFT} arrays ($F$ frequency bins and $T$ time frames) are represented by flattened 1D arrays, e.g., $\vect{s}\in\mathbb{C}^{FT}$.
\subsection{Speech generative modeling}\label{sec:diffusion_model}
Learning a speech generative prior using diffusion models involves smoothly injecting noise into training samples, via a diffusion process $\{\vect{s}_{t}\}_{t\in[0,1]}$, which transforms clean training data $\vect{s}_0=\vect{s}$ into Gaussian noise over time $t$. This can be described by the following \textit{forward} \ac{SDE} \cite{song2021scorebased, nortier2023unsupervised}
\begin{equation}\label{eqn:sde-fwd}
    \textrm{d}\vect{s}_t = \vect{f}(\vect{s}_t) \textrm{d}t + g(t) \textrm{d}\vect{w},
\end{equation}
where $\vect{w}$ denotes a standard Wiener process, the vector-valued $\vect{f}$ is the \textit{drift} coefficient term, the scalar function $g$ is the \textit{diffusion} coefficient, and $\textrm{d}t$ is an infinitesimal time-step. Here, $\vect{f}(\vect{s}_t)=-\gamma\vect{s}_t$, where $\gamma$ is a constant parameter, and $g(t)$ controls the variance of the stochastic noise. The \ac{SDE} in \eqref{eqn:sde-fwd} has the \textit{perturbation kernel} defined below, which allows one to directly sample $\vect{s}_t$ given $\vect{s}$
\begin{equation}\label{eqn:p0t-richter}
    p_{0t}(\vect{s}_t | \vect{s}) = \mathcal{N}_{\mathbb{C}}({\delta}_t \vect{s}, \sigma(t)^2\vect{I}),
\end{equation}
where ${\delta}_t = \textrm{e}^{-\gamma t}$, and the variance $\sigma(t)^2$ is determined from the \ac{SDE}. Under some light regularity conditions~\cite{anderson1982}, the noising process can be reverted through a \textit{reverse} \ac{SDE}: 
\begin{equation}\label{eqn:rev-sde}
    \textrm{d}\vect{s}_t = [-\vect{f}(\vect{s}_t)  + g(t)^2 \nabla_{\vect{s}_t} \log p_t(\vect{s}_t) ] \textrm{d}t + g(t) \textrm{d}\vect{\bar{w}},
\end{equation}
where $\vect{\bar{w}}$ is a standard Wiener process running backward in time. The term $\nabla_{\vect{s}_t} \log p_t(\vect{s}_t)$, known as the \textit{score function}, is intractable to compute directly. It is thus approximated by a time-dependent \ac{DNN}-based \textit{score model}, denoted as $\vect{S}_{\theta}(\vect{s}_{t}, {t})$. To learn $\theta$, the following problem is solved \cite{kingma2023understanding}:
\begin{equation}\label{eqn:train-obj}
    \theta^{*} = {\argmin_{\theta} \mathbb{E}_{t, \vect{s}, \bs{\zeta}, \vect{s}_t | \vect{s}}}
    {\Bigr[\| {\sigma(t)}{\vect{S}_{\theta}(\vect{s}_{t}, {t})}+ {\bs{\zeta}} \|_2^2 \Bigl]},
\end{equation}
where $\bs{\zeta}\sim\mathcal{N}_{\mathbb{C}}(\vect{0}, \vect{I})$ is a zero-mean complex-valued Gaussian noise. The reverse \ac{SDE} can then be numerically solved, e.g., using the \ac{PC} sampler \cite{song2021scorebased}, to sample from the data distribution.
\subsection{Unsupervised speech enhancement}
The additive noise is modeled as $\vect{n} \sim \mathcal{N}_{\mathbb{C}}(\bs{0}, \diag(\vect{m}_{\phi}))$, where $\vect{m}_{\phi}=\text{vec}(\vect{W}\vect{H})$, with $\vect{W}$, $\vect{H}$ being low-rank matrices with non-negative entries, and $\text{vec}(.)$ denoting the vectorization operator. Given the pre-trained speech prior with frozen parameters $\theta^{*}$, an iterative \ac{EM} method is performed to update $\phi$, where the M-step writes:
%is formulated below
\begin{equation}\label{eq:em_diff}
\phi \leftarrow\argmax_{\phi}~\E_{p_\phi(\vect{s}|\vect{x})}\lk\log p_\phi(\vect{x}| \vect{s})\rk.
\end{equation}
The above expectation is approximated using a Monte Carlo estimate, which involves sampling from the intractable posterior $p_{\phi}(\vect{s}|\vect{x}) \propto p_{\phi}(\vect{x}|\vect{s}) p(\vect{s})$ during the E-step. This approximation is implemented by substituting $\nabla_{\vect{s}_t} \log p_t(\vect{s}_t)$ in \eqref{eqn:rev-sde} with the following posterior-based score function:
\begin{equation}
    \nabla_{\vect{s}_t} \log p_{\phi}(\vect{s}_t|\vect{x}) = \nabla_{\vect{s}_t} \log p_{\phi}(\vect{x}|\vect{s}_t)+ \nabla_{\vect{s}_t} \log p_t(\vect{s}_t),
\end{equation}
where the time-dependent likelihood $p_{\phi}(\vect{x}|\vect{s}_t)$ is approximated with a noise-perturbed pseudo-likelihood \cite{nortier2023unsupervised} and the score function is replaced with $\vect{S}_{\theta^*}(\vect{s}_{t}, {t})$. Plugging the obtained clean speech estimate in \eqref{eq:em_diff}, $\vect{W}, \vect{H}$ are learned with multiplicative update rules. The EM steps are iteratively performed until convergence, typically requiring around five EM iterations for sufficient performance \cite{nortier2023unsupervised}.

\section{Diffusion-based unsupervised AVSE}\label{sec:proposed}
This section presents our proposed unsupervised \ac{AVSE} framework using diffusion models. We first develop an audio-visual speech prior model. Then, we propose a fast inference algorithm to estimate the clean speech. 
\subsection{Audio-visual speech generative model}\label{crossattn}
We model the conditional speech generative distribution $ p(\vect{s}|\vect{v})$, where $\vect{v}$ denotes a visual embedding associated with $\vect{s}$. Following the diffusion-based framework discussed in Section~\ref{sec:diffusion_model}, a conditional score network $\vect{S}_{\theta}(\vect{s}_{t}, \vect{v}, {t})$ is learned over clean AV speech data. The visual data denoted as $\vect{v} \in \mathbb{R}^{T_v \times p}$, where $T_v$ represents the number of video frames and $p$ indicates the embedding dimension, is incorporated into the score network as illustrated in Fig.~\ref{fig:av_score}.

%\subsubsection{AV-fusion in the score-network}
% \noindent\textbf{Cross attention for visual conditioning.} 
To integrate audio and visual features within the score network, a cross-attention mechanism is employed at each downsampling and upsampling stage of the U-Net-like architecture. Here, audio features serve as queries, while visual features are used as keys and values. More specifically, we denote the acoustic embedding features at the $i$\textsuperscript{th} layer of the score network as $e_{a, i} \in \mathbb{R}^{C_i \times F_i \times T_i}$, where $C_i$ represents the number of channels, and $F_i$ and $T_i$ indicate the embedding dimensions. Initially, the audio and visual features are projected into $d_i$-dimensional spaces for queries, keys, and values. This is followed by the computation of dot-product attention to produce a feature map of dimensions $C_i \times d_i \times T_i$. This feature map is then projected into an $F_i$-dimensional space, resulting in an intermediate audio-visual representation, $\widetilde{e}_{av, i} \in \mathbb{R}^{C_i \times F_i \times T_i}$. The final audio-visual representation at layer $i$, denoted $e_{av, i}$, is achieved by adding the original acoustic embedding $e_{a, i}$ to the Group Normalized intermediate representation: $e_{av, i} = e_{a, i} + \operatorname{GroupNorm}(\widetilde{e}_{av, i})$.

% More formally, we have:

% $z = \operatorname{softmax}\left(\frac{Q K^T}{\sqrt{d_i}}\right) V$

% where

% $\begin{cases} Q = e_{a, i}^{T} \cdot W_Q^{T} \text{ with } W_Q \in \mathbb{R}^{d_i \times F_i}
% \\ 
% K = \vect{v} \cdot W_K^{T}, \quad V = \vect{v} \cdot W_V^{T} \text{ with } W_K, W_V \in \mathbb{R}^{d_i \times p}
% \end{cases}$

% We then compute $\widetilde{e}_{av, i} = z \cdot W $ with $W \in \mathbb{R}^{F_i \times d_i} $ and the final audio-visual representation $e_{av, i}$ of layer $i$ is: $e_{av, i}= e_{a, i} + \operatorname{Groupnorm}(\widetilde{e}_{av, i})$

%\mostafa{Put a diagram of the av score model}

\subsection{Fast inference algorithm}
The UDiffSE framework requires multiple rounds of an iterative reverse diffusion process as the E-step to obtain an estimation of the clean speech by sampling from the posterior $p_\phi(\vect{s}|{\vect{x}})$. This method is computationally intensive. We introduce a significantly more efficient methodology named UDiffSE+, which requires only one round of reverse diffusion.

This new framework leverages a clean speech estimate at each iteration of the reverse diffusion process, eliminating the need for a complete reverse cycle. In this methodology, the noise parameters, \textit{i.e.,} $\vect{W}$ and $\vect{H}$, are updated following each reverse iteration based on the clean speech estimate obtained. This approach effectively employs an alternating maximization strategy, aimed at solving
% the following maximum a posteriori (MAP) problem
\begin{equation}\label{eq:fudiffse_prob}
\{ \phi^*, \vect{s}^*\}=\argmax_{\phi, \vect{s}}~\log p_\phi(\vect{x}| \vect{s}) + \log p(\vect{s}),
\end{equation}
by performing the following iterations
\begin{subequations}
    \begin{empheq}[left={\empheqlbrace}]{align}
        \vect{s}_{0,k+1} &= \argmax_{\vect{s}}~\log p_{\phi_{k}}(\vect{x}| \vect{s}) + \log p(\vect{s}), \label{eq:s_update}\\
        \phi_{k+1} &= \argmax_{\phi}~\log p_\phi(\vect{x}| \vect{s}_{0,k+1}). \label{eq:phi_update}
    \end{empheq}
\end{subequations}
Problem~\eqref{eq:s_update}, which is maximum a posteriori (MAP) estimation, can be solved by performing one reverse iteration, as done in UDiffSE, in which case, the iteration index $k$ is replaced with a time discretization, denoted $\tau$. To obtain an estimate of $\vect{s}$ at iteration $\tau$ of the reverse diffusion process, denoted $\hat{\mathbf{s}}_{0,\tau}$, we leverage Tweedie's formula \cite{efron2011tweedie}, that is
\begin{equation}
    \hat{\vect{s}}_{0,\tau}=
    \E_{p_{\tau 0}(\vect{s}_0|\vect{s}_\tau)}[\vect{s}_0]\approx\frac{\vect{s}_\tau + \sigma_{\tau}^2 \vect{S}_{\theta^*}(\vect{s}_\tau, \tau)}{\delta_{\tau}}.
\end{equation}
Plugging this into \eqref{eq:phi_update}, the parameters are updated by a single multiplicative update iteration. The overall (audio-only/AV) UDiffSE+ algorithm is described in Algorithm~\ref{alg:diffuse}. By excluding lines 12 and 13, the algorithm simplifies to the E-step of UDiffSE. The highlighted box corresponds to the operations that ensure observation ($\vect{x}$) consistency, without which the algorithm reduces to prior sampling. 

\begin{algorithm}[t!]
  \scriptsize
\caption{A{\textcolor{blue}{V}}-UDiffSE+}\label{alg:diffuse}
\begin{algorithmic}[1]
\Require $\vect{x}, \textcolor{blue}{\vect{v}}, N, \lambda, r (\mbox{\small signal-to-noise ratio})$
    \State ${\vect{s}}_1 \sim \mathcal{N}_{\mathbb{C}}(\vect{x}, \vect{I}), \Delta \tau \gets \frac{1}{N}$
    \For{$i=N,\ldots, 1$}
        \State $\tau\gets\frac{i}{N}$
        \State $\epsilon_\tau \gets (\sigma_\tau\cdot r)^2$
        \State $\bs{\zeta}_c\sim\normalc$ \hspace{3mm} \Comment{\textit{(Corrector)}}
        \State ${\vect{s}}_{\tau} \gets \vect{s}_\tau + \epsilon_\tau {\vect{S}_{\theta^*}(\vect{s}_\tau, \textcolor{blue}{\vect{v}}, {\tau})} + \sqrt{2\epsilon_\tau}\bs{\zeta}_c$        
        % \Statex 
        \State $\bs{\zeta}_p\sim\normalc$ \hspace{3mm} \Comment{\textit{(Predictor)}}
        \State ${\vect{s}}_{\tau} \gets {\vect{s}}_{\tau} - \vect{f}_{\tau}\Delta{\tau} + g_{\tau}^2 \vect{S}_{\theta^*}({\vect{s}}_{\tau}, \textcolor{blue}{\vect{v}}, \tau) \Delta{\tau}+  g_{\tau}\sqrt{\Delta{\tau}}\bs{\zeta}_p $
        
        \tikzmk{A}\If{$i \equiv 0 \pmod{2}$} \hspace{3mm} \Comment{\textit{(Posterior)}}
            \State $\nabla_{{\vect{s}}_{\tau}} \log \tilde{p}_\phi (\vect{x}|{\vect{s}}_{\tau}) \gets\dfrac{1}{{\delta}_\tau}\Big[\dfrac{{\sigma^2_\tau}}{{{\delta}^2_\tau}} \vect{I} + \diag(\bs{v}_{\phi}) \Big]^{-1}  
             (\vect{x}-\dfrac{{\vect{s}}_{\tau} }{\delta_\tau}) $
            \State $\vect{s}_{\tau} \gets {\vect{s}}_{\tau}  + {\lambda}{g_{\tau}^2} {\nabla_{\vect{s}_\tau} \log \tilde{p}_\phi (\vect{x}|\vect{s}_\tau)} \Delta{\tau}$
            \State {$\hat{\mathbf{s}}_{0,\tau}
    = \delta_{\tau}^{-1}\Big(\mathbf{s}_\tau + \sigma^2_\tau \vect{S}_{\theta^*}(\mathbf{s}_\tau, \textcolor{blue}{\vect{v}}, \tau)\Big)$} \hspace{3mm} \Comment{\textit{(Estimate of $\vect{s}_0$)}}
            \State {$\phi \gets \argmax_{\phi}\log p_\phi(\vect{x}| \hat{\mathbf{s}}_{0,\tau})$} \hspace{3mm} \Comment{\textit{(Parameters update)}}
        \EndIf
        \tikzmk{B}\boxit{myblue}
    \EndFor
    \State \Return $\hat{\vect{s}}={\vect{s}}_{0}$
\end{algorithmic}
\end{algorithm}

\begin{table*}[!t]
\centering
\caption{Speech enhancement results in matched (TCD speech + DEMAND noise) and mismatched (LRS3 speech + NTCD noise) conditions. AO and AV: audio-only and audio-visual models, respectively. \textup{EM}: number of EM iterations. \textup{RTF} (real-time factor): average time needed to process one second of speech. Values: mean $\pm$ standard error, \textbf{best value}, \underline{next best} per column.} %Bold and underlined values are the best and next-best values in each column.} 
%\caption{Speech enhancement results in matched (TCD speech + DEMAND noise) and mismatched (LRS3 speech + NTCD noise) conditions. AO and AV: audio-only and audio-visual models, respectively. \textup{EM}: number of EM iterations. \textup{RTF} (real-time factor): average time needed to process one second of speech. Values indicate mean and standard error.}
\resizebox{0.95\linewidth}{!}{%
\begin{tabular}{lcccccrccr}
\toprule
                                 &    & & &\multicolumn{3}{c}{{TCD speech + DEMAND  noise}} & \multicolumn{3}{c}{{LRS3 speech + NTCD noise}} \\
                                  \cmidrule(lr){5-7} \cmidrule(lr){8-10} 
\multicolumn{1}{l}{Method}                       & \# Params (M) & EM &RTF $\downarrow$       & SI-SDR $\uparrow$     & PESQ $\uparrow$      & \multicolumn{1}{c}{ESTOI$\uparrow$}      &  SI-SDR $\uparrow$    & PESQ $\uparrow$    & \multicolumn{1}{c}{ESTOI$\uparrow$} \\
\midrule
\multicolumn{1}{l}{Input} &  &  &   & 0.00\ {\scriptsize $\pm$0.17} & 2.83\ {\scriptsize $\pm$0.02} & 0.70\ {\scriptsize $\pm$0.01} & 0.03\ {\scriptsize $\pm$0.14} & 2.10\ {\scriptsize $\pm$0.02} & 0.58\ {\scriptsize $\pm$0.01} \\ \midrule
\multicolumn{1}{l}{UDiffSE~\cite{nortier2023unsupervised}}  &27.7 &~~5 &10.53   &11.69\ {\scriptsize $\pm$0.27} & 3.08\ {\scriptsize $\pm$0.02} & 0.77\ {\scriptsize $\pm$0.01} & \underline{4.72\ {\scriptsize $\pm$0.16}} & \underline{2.38\ {\scriptsize $\pm$0.02}} & \underline{0.62\ {\scriptsize $\pm$0.01}} \\ 
\multicolumn{1}{l}{FlowAVSE~\cite{jung2024flowavse}}  &60.2 &~~~~ &\textbf{0.03}           &  \textbf{17.83\ {\scriptsize $\pm$0.18}} & 3.18\ {\scriptsize $\pm$0.02} & \textbf{0.82\ {\scriptsize $\pm$0.00}} & 3.12\ {\scriptsize $\pm$0.23} & 1.49\ {\scriptsize $\pm$0.02} & 0.53\ {\scriptsize $\pm$0.01} \\
\midrule
\multicolumn{1}{l}{AV-UDiffSE}  &29.4 &~~5 &13.92 &\underline{13.70\ {\scriptsize $\pm$0.23}} & 3.18\ {\scriptsize $\pm$0.02} & \underline{0.79\ {\scriptsize $\pm$0.01}} & \textbf{5.60\ {\scriptsize $\pm$0.15}} & \textbf{2.48\ {\scriptsize $\pm$0.02}} & \textbf{0.64\ {\scriptsize $\pm$0.01}} \\ 
\multicolumn{1}{l}{AO-UDiffSE+}  &27.7 &~~1 &\underline{1.98} &  8.99\ {\scriptsize $\pm$0.21} & \underline{3.19\ {\scriptsize $\pm$0.02}} & 0.72\ {\scriptsize $\pm$0.01} & 2.95\ {\scriptsize $\pm$0.15} & 2.31\ {\scriptsize $\pm$0.02} & 0.59\ {\scriptsize $\pm$0.01} \\
\multicolumn{1}{l}{AV-UDiffSE+}  &29.4 &~~1 &{2.68} & 10.21\ {\scriptsize $\pm$0.15} & \textbf{3.26\ {\scriptsize $\pm$0.02}} & 0.74\ {\scriptsize $\pm$0.00} & 3.67{\scriptsize $\pm$0.13} & 2.42\ {\scriptsize $\pm$0.02} & 0.61\ {\scriptsize $\pm$0.01}  \\
\bottomrule
\end{tabular}
}
\label{tab:new_se_results}
\end{table*}

\section{Experiments}\label{sec:exp}

% \subsection{Baseline}
\noindent\textbf{Baselines.} We compare the performance of our proposed method against two frameworks: the audio-only UDiffSE \cite{nortier2023unsupervised} %(\ref{sec:udiffse})
and the supervised-generative FlowAVSE model \cite{jung2024flowavse}.
% \subsection{Dataset}

\noindent\textbf{Dataset.} The TCD-TIMIT corpus \cite{harte2015tcd} was employed for training. It comprises AV speech data from 56 English-speaking individuals with Irish accents, distributed among 39 for training, 8 for validation, and 9 for testing. The dataset features 98 distinct sentences, each approximately 5 seconds in duration and sampled at 16 kHz, totaling around 8 hours of data. Additionally, each spoken utterance is accompanied by a corresponding video, recording the speaker from a frontal perspective at a frame rate of 30 fps. We downsampled the videos to 25 fps and the lip regions of interest are extracted as 88$\times$88 grayscale following \cite{ ma2022visual, yemini2024lipvoicer}. Training the supervised baseline model requires noisy speech counterparts. As such, we consider mixing TCD-TIMIT clean speech with DKITCHEN, OMEETING, PRESTO, PSTATION, NPARK noises from the DEMAND dataset \cite{thiemann2013diverse} at \{-10, 0, 10\}dB.   %For each video, the lip regions of interest are extracted as 67$\times$67 grayscale images. 

%For evaluating
To evaluate SE performance, we define two scenarios: \textit{matched} and \textit{mismatched}, based on the origin of the test clean speech signals or noises. In the matched scenario, noisy speech is constructed by mixing clean speech from the TCD-TIMIT test set with noises from the DEMAND dataset (TMETRO, OOFFICE, TBUS, STRAFFIC, SPSQUARE). For each type of noise and SNR level, we randomly selected 10 utterances from each test speaker, totaling 900 evaluation utterances. In the mismatched scenario, we use the test set from the LRS3 dataset \cite{Afouras2018LRS3TEDAL}, consisting of 1321 clean speech files. These files are mixed randomly with one of the noise types from the NTCD-TIMIT dataset (Living Room, White, Car, Babble, Cafe) \cite{abdelaziz2017ntcd}. For both scenarios, the SNR levels are \{-5, 5\}dB.

% we used noisy speech samples from the NTCD-TIMIT corpus \cite{abdelaziz2017ntcd}. This corpus includes test speech data from TCD-TIMIT, mixed with six noise types—\textit{\ac{LR}}, \textit{White}, \textit{Cafe}, \textit{Car}, \textit{Babble}, and \textit{Street}—at various \ac{SNR} levels. For each type and level of noise, 5 utterances were randomly selected from each test speaker's recordings, totaling 1350 utterances for evaluation.
% \subsection{Evaluation metrics}

\noindent\textbf{Evaluation metrics.} We use standard SE metrics: the scale-invariant signal-to-distortion ratio (SI-SDR) in dB \cite{le2019sdr}, the extended short-term objective intelligibility (ESTOI) measure \cite{jensen2016algorithm}, varying between $[0,1]$, and the perceptual evaluation of speech quality (PESQ) score \cite{Rix2001pesq}, with a range of $[-0.5,4.5]$. For these metrics, higher values indicate better performance.

%===== or choose the table =======

\noindent\textbf{Model architecture.} The base architecture in this paper, NCSN++M, is a lighter U-Net-like version of the original NCSN++ network \cite{richter2023speech}. In its audio-only configuration, NCSN++M comprises 27.8 million parameters. Integration of video data through cross-attention modules, each featuring a single attention head as detailed in Subsection~\ref{crossattn}, results in a 6.13\% increase in the total number of parameters. Additionally, the FlowAVSE baseline employs the NCSN++M network in both stages, totaling 60.2 million parameters. All in all, we trained from scratch the score networks for the audio-only and AV settings and the FlowAVSE networks.

% In the AV setting, the same architecture as in the audio-only setup is retained and augmented by adding cross-attention modules featuring a single attention head layer.

%\mostafa{Discuss types of visual features}
% \subsection{Pretrained visual features}
\noindent\textbf{Pretrained visual features.} Similar to prior works  \cite{chern2023audio, chou2024av2wav, richter2023audio}, we extracted features from lip videos using the AV-HuBERT model \cite{shi2022learning}, specifically utilizing its version fine-tuned for visual speech recognition. We maintained this visual encoder in a frozen state throughout our experiments. It processes silent grayscale videos and outputs visual features $\vect{v}$ of size $(T_v, p)$, with $p=768$, derived from the model’s final layer. For visual feature extraction in the FlowAVSE baseline, we followed the procedures outlined in the original paper \cite{jung2024flowavse}.

% \noindent\textbf{Pretrained visual features.} Similarly to \cite{chern2023audio, chou2024av2wav, richter2023audio}, we extracted features from lips video by leveraging the self-supervised audio-visual representation learning model AV-HuBERT \cite{shi2022learning}. Specifically, we used its fine-tuned version for visual speech recognition and kept it frozen. This visual encoder processes silent grayscale videos shaped $(C, T_v, H, W)$, where $C=1$ specifies the channel, $T_v$ counts the video frames, and $H$ and $W$ are the image's height and width, respectively. As output, a video embedding $\vect{v}$ of size $(T_v, p)$ (with embedding size $p=768$) from the final layer is obtained. For visual feature extraction in the FlowAVSE baseline, we followed the processing described in their paper \cite{jung2024flowavse}.
% For both models, we use the feature obtained at the last layer.
% AV-HuBERT is a self-supervised AV speech representation learning model whose architecture mainly consists of successive transformer encoder layers.
% We do not finetune these visual feature extractors when training the score network.

% \subsection{Hyperparameters setting.} 
%We broadly reused the hyperparameters of  \cite{nortier2023unsupervised}. As such,
\noindent\textbf{Hyperparameters setting.} We trained the diffusion models over 200 epochs using the Adam optimizer %\cite{kingma2015adam} 
at a learning rate of 0.0001. For the STFT, a Hann window of 510 length and a hop length of 128 were used, yielding 256 frequency bins. Each speech waveform was randomly trimmed to 2.04 seconds, corresponding to $T=256$ STFT time frames, to standardize input sizes for batching. The hyperparameters for the \ac{SDE} and inference algorithm were aligned with those in \cite{nortier2023unsupervised}.% (5 EM iterations). 

% \subsection{Results}
%We report in table \ref{tab:se_results},
\noindent\textbf{Results.} Table \ref{tab:new_se_results} presents the SE performance metrics for audio-only (AO) and audio-visual (AV) algorithms under different conditions. In the matched scenario, the supervised baseline generally outperforms the unsupervised models in all metrics, except for AV-UDiffSE+ in PESQ, which aligns with the typical strengths of supervised methods in matched settings. Conversely, in the mismatched scenario, all methods, including the audio-only unsupervised models, surpass the FlowAVSE baseline by at least 0.55 dB in SI-SDR, 0.82 in PESQ, and 0.06 in ESTOI, except for AO-UDiffSE+ in SI-SDR. This indicates a potential for generalization in diffusion-based unsupervised SE frameworks, as also suggested by \cite{nortier2023unsupervised}. It is important to highlight that like supervised methods, unsupervised models also offer the flexibility to incorporate the visual modality. As expected, incorporating lip movements into the prior diffusion models (AV-UDiffSE+, AV-UDiffSE) enhances noise removal, speech quality, and intelligibility. %compared to using audio-only (UDiffSE+, UDiffSE). %This improvement has been confirmed through informal listening tests. 
%For listening examples, please visit our project page\footnote{\url{https://jeaneudesayilo.github.io/fast_UdiffSE}}.

% \noindent\textit{Inference speed vs Performance.} 
We assessed the inference speed of the different %various 
algorithms on an Nvidia A100-SXM4 (40GB), maintaining consistent conditions across all tests. Comparisons between UDiffSE and our proposed UDiffSE+ frameworks (in both audio-only and AV settings) revealed that UDiffSE, while superior in performance, requires significantly more processing time, as indicated by the real-time factor (RTF); it is approximately 5 times slower than UDiffSE+. Thus, UDiffSE+ offers a favorable compromise between inference speed and performance.

\section{Conclusion}\label{sec:conc}
In this paper, we present a diffusion-based, unsupervised audio-visual speech enhancement (AVSE) framework, which leverages a diffusion model to simulate clean speech distribution, conditioned on visual cues from lip movements. The pre-trained diffusion model is integrated with an NMF-based noise model through an iterative process of reverse diffusion steps to estimate speech. Our experiments show that the proposed AVSE framework consistently outperforms its audio-only counterpart and offers better generalization than a recent supervised-generative approach \cite{jung2024flowavse}.
%can have a slight edge in generalization performance compared to a recent supervised-generative approach \cite{jung2024flowavse}.
Moreover, compared to the previous inference method \cite{nortier2023unsupervised}, our algorithm achieves a better trade-off between performance and runtime. Future work includes a comprehensive subjective performance assessment.

\bibliographystyle{IEEEtran_abbrev}
\bibliography{mybib}

\end{document}